\documentclass[conference]{IEEEtran}
\IEEEoverridecommandlockouts

\newcommand\blfootnote[1]{%
  \begingroup
  \renewcommand\thefootnote{}\footnote{#1}%
  \addtocounter{footnote}{-1}%
  \endgroup
}

\usepackage{cite}
\usepackage{adjustbox}
\usepackage{amsmath,amssymb,amsfonts}
\usepackage{graphicx}
\usepackage{textcomp}
\usepackage{xcolor}
\usepackage{multicol}
\usepackage{multirow}
\usepackage{subcaption}
\usepackage{enumitem}
\usepackage{balance}
\usepackage{booktabs}
\usepackage{array}
\newcolumntype{x}[1]{>{\centering\arraybackslash\hspace{0pt}}p{#1}}

\usepackage{pst-all}
\usepackage{pstricks-add}
\usepackage{float}

\usepackage{pgfplots}

\usepackage{tikz}
\usepackage{tikzscale}
\usepackage{transparent}

\usepackage{algorithm}
\usepackage[noend]{algpseudocode}

\usepackage{comment}

\usepackage{blindtext}

\definecolor{purplelight}{HTML}{ead1dc}
\definecolor{purpledark}{HTML}{a64d79}

\definecolor{purple}{HTML}{733d87}

\definecolor{color1}{HTML}{e6194B}

\definecolor{color2}{HTML}{3cb44b}

\definecolor{color3}{HTML}{ffe119}

\definecolor{color4}{HTML}{4363d8}

\definecolor{color5}{HTML}{f58231}

\definecolor{color6}{HTML}{911eb4}

\definecolor{color7}{HTML}{42d4f4}

\definecolor{color8}{HTML}{f032e6}

\definecolor{color9}{HTML}{bfef45}

\definecolor{color10}{HTML}{fabebe}

\definecolor{color11}{HTML}{469990
}
\definecolor{color12}{HTML}{e6beff}

\definecolor{color13}{HTML}{9A6324}

\definecolor{color14}{HTML}{a9a9a9}

\definecolor{color15}{HTML}{800000}

\def\BibTeX{{\rm B\kern-.05em{\sc i\kern-.025em b}\kern-.08em
    T\kern-.1667em\lower.7ex\hbox{E}\kern-.125emX}}
\begin{document}


\title{Computational Resource Allocation for Edge Computing in Social Internet-of-Things\vspace{-0.2cm}}
\author{
\IEEEauthorblockN{Abdullah Khanfor$^1$,
  Raby Hamadi$^1$,
  Hakim Ghazzai$^1$,
  Ye Yang$^1$, Mohammad Rafiqul Haider$^2$
  and Yehia Massoud$^1$\\
\IEEEauthorblockA{$^1$School of Systems \& Enterprises, Stevens Institute of Technology, Hoboken, NJ, USA}
\IEEEauthorblockA{$^2$University of Alabama at Birmingham, AL, USA}
}\vspace{-0.6cm}}

\maketitle

\begin{abstract}
The heterogeneity of the Internet-of-things (IoT) network can be exploited as a dynamic computational resource environment for many devices lacking computational capabilities. A smart mechanism for allocating edge and mobile computers to match the need of devices requesting external computational resources is developed. In this paper, we employ the concept of Social IoT and machine learning to downgrade the complexity of allocating appropriate edge computers. We propose a framework that detects different communities of devices in SIoT enclosing trustworthy peers having strong social relations. Afterwards, we train a machine learning algorithm, considering multiple computational and non-computational features of the requester as well as the edge computers, to predict the total time needed to process the required task by the potential candidates belonging to the same community of the requester. By applying it to a real-world data set, we observe that the proposed framework provides encouraging results for mobile computer allocation.
\end{abstract}

\blfootnote{\hrule
\vspace{0.2cm} This paper is accepted for publication in 63rd IEEE International Midwest Symposium on Circuits and Systems (MWSCAS'20), Springfield, MA, USA, Aug. 2020. \newline \textcopyright~2020 IEEE. Personal use of this material is permitted. Permission from IEEE must be obtained for all other uses, in any current or future media, including reprinting/republishing this material for advertising or promotional purposes, creating new collective works, for resale or redistribution to servers or lists, or reuse of any copyrighted component of this work in other works.}%

\begin{IEEEkeywords}
Internet of Things (IoT), community detection, machine learning, edge computing.
\end{IEEEkeywords}
\vspace{-0.3cm}
\section{Introduction}
\vspace{-0.1cm}
Efficiently utilizing the vast network of devices in the Internet-of-Things (IoT) system to leverage computational resources can be beneficial for edge computing services. The edge computing affirms to bring the available distributed, but yet closer resources, to the devices requesting external computational and/or storage capabilities~\cite{cicirelli2017edge}. In many cases, the resource sharing and computing capabilities are indispensable in the IoT system, considering many of terminal devices such as sensors, mobile phones, actuators, and personal computers that may lack these resources to accomplish specific tasks~\cite{yu2017survey}. Devices, in the neighborhood, can share their available computational resources to process tasks for the profit of their peers and help relieve the load of cloud and edge servers~\cite{zhang2018mobile}. Service discovery in IoT platforms can be used to seek for edge computing providers. Despite that, searching in the vast network to find suitable devices remains one of the major challenges in ubiquitous IoT networks, 
especially when devices are heterogeneous and require a variety of services with different levels of storage and computational capabilities at irregular time instants.


To achieve a productive search of mobile edge computers in the large-scale IoT, it is required to find available, trustworthy, and reliable devices that can potentially handle the targeted computational tasks. The emerging concept of Social IoT (SIoT) can help achieve these goals using the social relations built among the devices, \textit{aka} social objects~\cite{marche2018dataset}. 
These social relations transform the IoT system into a social network of devices or ``friends" having common characteristics and criteria, which can raise the level of security and trustworthiness in such diverse networks\cite{khanfor2020trustworthy}.

A number of studies have been proposed to address the implementation of edge computing in the IoT context. Alsaffar et al.~\cite{alsaffar2016architecture} proposed a high-level architecture to enable the IoT devices for service delegation and resource allocation to the nearest computing fog. Another study proposed profiling of the mobile IoT devices and managing the computation offloading with bandwidth constraint by considering the power consumption limitation \cite{samie2016computation}. Renner et al.~\cite{renner2016towards} proposed the concept of containerization to the IoT devices to cluster them and utilize their computing and storage resources.
Besides, the establishment of different social connections in SIoT is one of the thrust research areas, which can affect the structure of various clusters of devices and the way of differentiating between them. Exploiting these connections can aid in locating trustworthy devices~\cite{khanfor2019application}. For example, the friendship level among the devices' owners can serve to provide a level of access or authentication between the devices.
\begin{figure*}[t]
\centerline{\includegraphics[width=\textwidth]{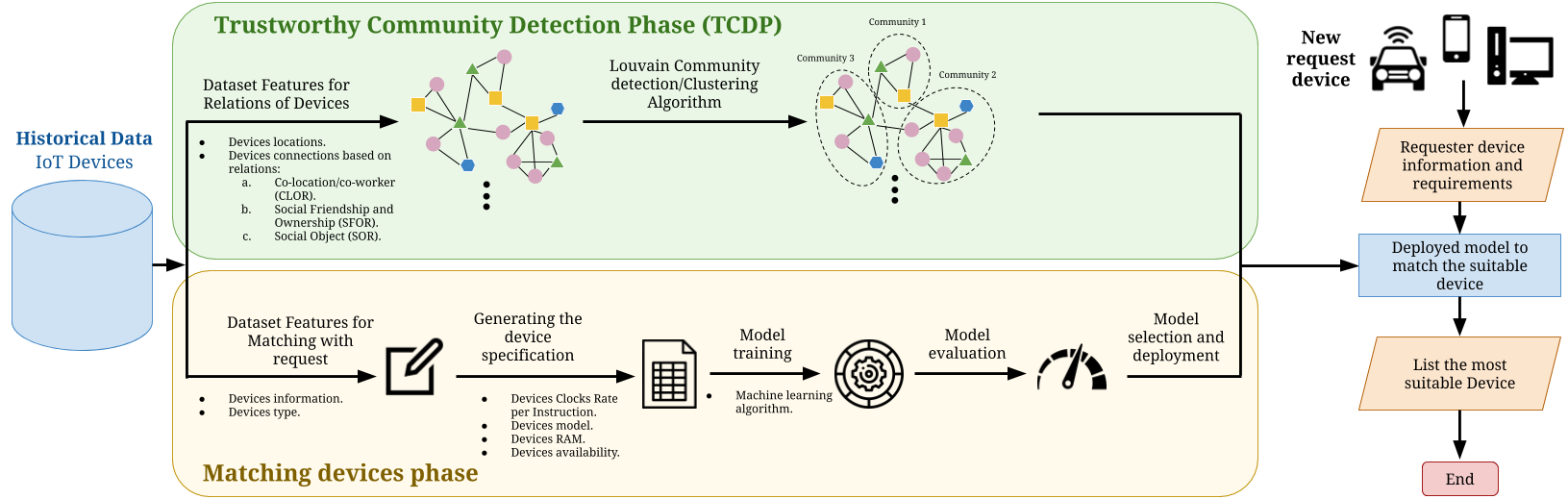}}
\caption{Proposed framework to identify communities and match devices for computational request in SIoT.}
\vspace{-0.4cm}
\label{fig:framework}
\end{figure*}

In this paper, we propose to develop a generic framework aiming at selecting appropriate edge computers that can handle the computational task of its peer in a trustworthy and rapid manner. To this end, we proceed with the two following phases: (i) implement a community detection algorithm, namely the Louvain algorithm, to divide the IoT network into multiple groups of IoT devices sharing strong social relations, then, (ii) run a machine learner, trained on previous resource sharing activities as well as the computational and non-computational features of the devices and the task to be executed itself, to predict the time to process the task. Starting from a request of an IoT device, the proposed framework outputs an edge computer socially connected to the requester and capable of rapidly executing the task. Our selected results, applied to a real-world data set, illustrate the effective operation of the framework. 
\vspace{-0.2cm}
\section{Proposed Framework}
\vspace{-0.1cm}
The proposed framework, presented in Fig.~\ref{fig:framework} is mainly composed of the two following phases. The first phase, called Trustworthy Community Detection Phase (TCDP), aims to detect communities regrouping IoT devices having strong social relations. The objective of this phase is to shrink the search space of the suitable devices and limit it to trustworthy devices for the requester of the computation task, e.g., closer devices in terms of location and ownership. The second phase, called the Matching Device Phase (MDP), aims to predict the time needed by the potential edge computer candidates, belonging to the same community of the requester, to accomplish the task and determine the one, that has necessary capabilities to provide the fastest service time.
\subsection{Trustworthy Community Detection Phase}
\vspace{-0.1cm}
Generally, different social relations between IoT devices can be created based on various criteria. We list below three different relations used in this study to establish trustworthy social among the IoT devices:

$\bullet$ \textit{Co-location/co-work based relation (CLOR):} The geographical locations of the devices can be used to define a certain relation reflecting the fact that two devices, static or mobile, are co-located in a given area at a certain instant of time. In some applications, it might be imposed, e.g., by the device owner, that a device collaborate only with other IoT devices in the neighborhood. By setting a defined threshold for the distance between the devices, we can specify if these devices belong to a common cluster and hence, establish relations between them based on their separating distances.

$\bullet$ \textit{Social friendship and ownership relation (SFOR):}
This relationship can stem from two principles. First, each two devices owned by the same owner are considered having full trustworthy relationship. Second, using the social network of the devices' owners, SFOR relations can be set between the devices according to the friendship levels of their owners. 


$\bullet$ \textit{Social object relationship (SOR):} This relationship can be built if there is a continuous collaboration or sporadic form of collaboration between two IoT devices. Owner policies are setting this relation between the devices. For example, if two device exchange data for a certain period, then a SOR can be established between them. Based on the historical activity of exposure between the devices and the owners' policies, SOR relations are created.

These three social IoT relations are used to construct three undirected and weighted graphs, respectively. The nodes in the graphs represent the IoT devices and the edges between them imply the different social IoT relationships. Moreover, the values of the edges' weights describe the strength of the connection between the nodes. For example, in SFOR, if the owners of the devices are direct friends or friends of a friend, then the weight of the link is calculated to reflect this level of friendship. It is worth to note that the nodes do not have a self-loop edge since none of the social relations requires to imply that in the graph.

After establishing the social relationships between all SIoT devices in the area of interest, we propose to apply the Louvain method to identify the different communities based on these relations~\cite{blondel2008fast}. This allows the division of the IoT network into virtual groups based on these social relationships and determine the groups of devices having more chance to collaborate with each other in a safe and trustworthy manner. We advocate the use of the Louvain algorithm mainly because of its running time of $O(n \log n)$ to identify the communities.
The final set of devices, denoted by $\mathcal S_r$, corresponds to the intersection of CLOR, SFOR, and SOR communities enclosing the requester device. In this way, we have shrunk the search space to find edge computers from all the IoT devices of the network to a limited space encompassing trustworthy devices to the requester. The union of these communities, especially of SFOR and SOR can be also considered for a relaxed search space. In this sense, a device having an SFOR or SOR relationship can be considered as a trustful edge computer.
\vspace{-0.2cm}
\subsection{Edge Computer Selection} \label{matching}
\vspace{-0.1cm}
Not all of the devices obtained from the TCDP are capable of handling the requested computational tasks. In this phase, the proposed framework provides an intelligent process to determine the suitable available edge computer. The MDP is performed with regard to the features of the requester, edge computer candidates, and the computational task itself. To this end, we propose to employ machine learning techniques to predict the time needed by an edge computer, belonging to the set $\mathcal S_r$, to accomplish a given task based on the historical computational sharing activities of the IoT system. Based on the results of the previous sharing experiences and the features of the IoT devices (requester and edge computer), we train a machine learning algorithm to estimate the time required to accomplish the task of all members of the trustworthy clusters to select the edge computer providing the lowest total response time while considering the average round-trip propagation time of the task's messages as well as its processing time.

\subsubsection{IoT Devices Data set}

The IoT data set used in this paper includes features describing the computational capabilities of the devices. 
The IoT network can be seen as a network composed by two non-disjoint sets of devices: edge computers and requester devices.

\paragraph{Requester Devices}
They are IoT devices seeking for external computational resources to handle a specific task. In order to submit a task, a requester defines features to be used in finding suitable edge computers to handle the task such as:\\
$\bullet$ \textit{Location}: Geographical coordinates (longitude, latitude).\\
$\bullet$ \textit{Type}: Categorical variable indicating the device's type, e.g, vehicle, PC, smartphone, smartwatch, etc.\\
$\bullet$ \textit{IC}: The number of instructions within the task.\\
$\bullet$ \textit{M}: The size of the task's message.
\paragraph{Edge Computers}
These devices are also IoT devices part of the same network and are supposed to handle the tasks submitted by their peers. The edge computers are mainly distinguished with features indicating their computational power levels such as:\\
$\bullet$ \textit{CPI}: Clocks per Instruction. \\
$\bullet$ \textit{R}: The clock rate of the processor installed on the device. \\
$\bullet$ \textit{RAM}: Indicates the installed RAM capacity.\\
$\bullet$ \textit{Type}: Categorical variable indicating the device's type.\\
$\bullet$ \textit{Mode}: Indicates if the device is private or public.\\
$\bullet$ \textit{Location}: Geographical coordinates of the device.\\
$\bullet$ \textit{Mobility}: Indicates if the device is static or mobile.\\
$\bullet$ \textit{Availability}: Numerical value indicating the availability of the computational resources of the device ranged between 0\% and 100\%. This feature is set to 0\% for devices with no computational power within the network.


All these parameters are included in the IoT data set in addition to other features such as the devices' brand, the devices' owner, the number of cores, and the manufacturer of the processor installed on each device. We also consider another categorical feature indicating the communication technology used by each pair of devices based on their separating distance, e.g., cellular or device-to-device (D2D) communications.

\subsubsection{Model Building}
The estimation of the time needed to accomplish a task can be seen as a regression problem. 
We use machine learning techniques to predict the total response time of each potential edge computer within the optimized search zone, i.e., output of the TCDP. The delegated edge computer is the IoT device expected to provide less total response time while executing the given task. We investigate three regression models to predict the expected total response time between the two entities. Random Forest (RF) is a commonly used supervised machine learning algorithm proposed by Breiman~\cite{10.1023/A:1010933404324}. Its applications are various since it can be used in both classification and regression problems. Gradient Boosting Regressor (GBR) combines multiple weak base models to form a committee whose performance outperforms the base models. A new weak based-model is added to the committee and trained with respect to the loss of the whole learnt ensemble on each iteration of the training process~\cite{10.5555/1162264}. Decision Tree (DT) is one of the most popular supervised learning algorithms used for both classification and regression tasks. This algorithm is able to identify the attributes with higher information gain than the rest of the attributes~\cite{article}. In the next section, we apply these algorithms to a real-world data set and evaluate their performances for different metrics.

\vspace{-0.2cm}
\section{Results \& Discussions}\label{sec:results}
We use an IoT data set of Santander, Spain, from Marche et al. \cite{marche2018dataset} to analyze our proposed framework in a real-world case study that includes 2568 heterogeneous devices. 
Following that, we establish the links between the devices by incorporating the different social relations descriptions for SIoT of SFOR and SOR.


\paragraph{SIoT Community Detection}For the SFOR relation, we generate a social network between the owners using Watts–Strogatz generator \cite{watts1998collective} since we do not have the social network between the owners. From the social network of owners, we compute the weights of friends' owner devices such that every direct friend in the social network have an edge of 0.5 between their devices in the SIoT network. For a friend of friend owners, the weight of the edges is degreed by calculating the division of the number of friends to reach the owner of the device. Hence, the weight between the nodes in the SFOR represents the social relation strength between the owners. Finally, for SOR relation, we get Marche et al.~\cite{marche2018dataset} simulation of the connection of devices. A SOR relationship is assumed to be established between two devices if they have met three times or more within a duration of 30 minutes. The simulation is conducted for ten days.

\begin{figure*}[t]
    \centering
    \begin{subfigure}{0.73\columnwidth}
        \includegraphics[width=\textwidth]{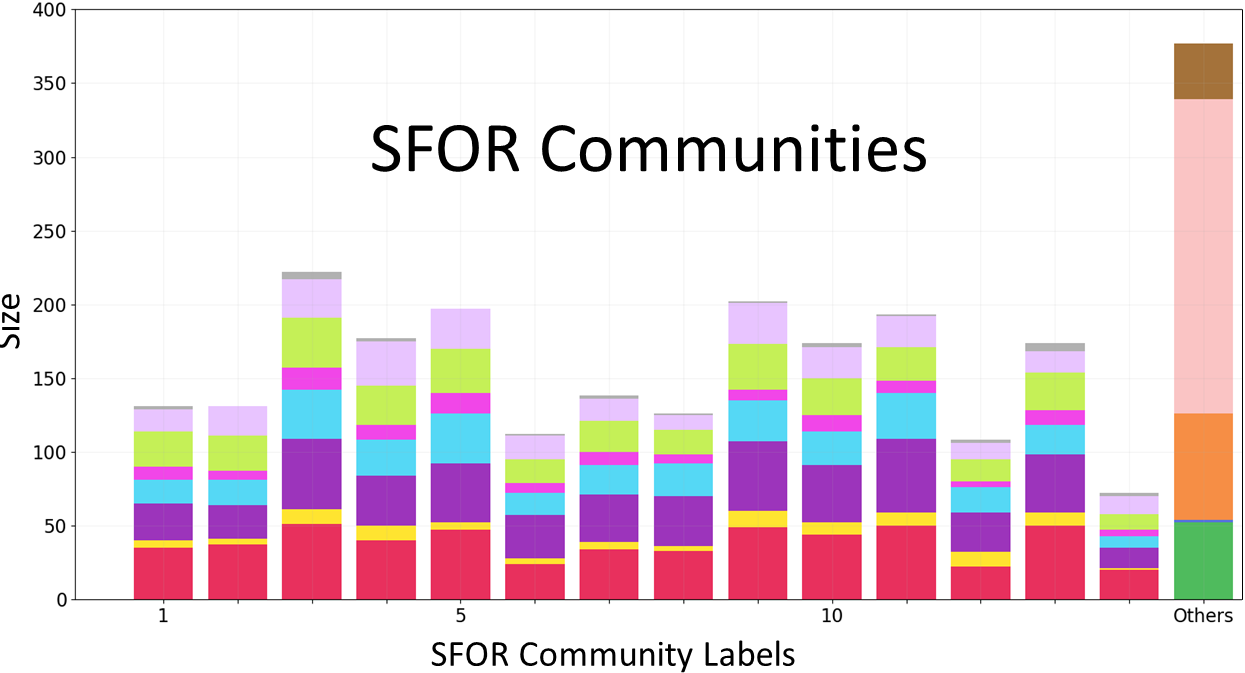}
        \label{fig:SFOR} 
    \end{subfigure}
    \begin{subfigure}{0.73\columnwidth}
    \includegraphics[width=\textwidth]{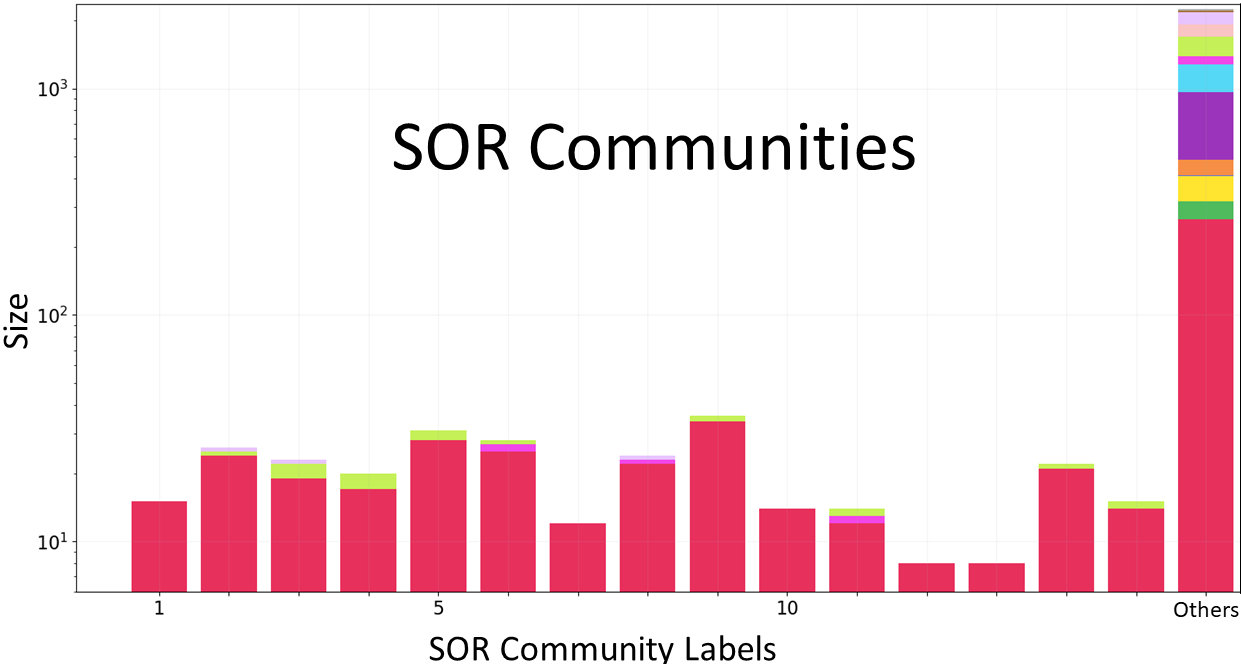}
        \label{fig:SOR} 
    \end{subfigure}
    \begin{tabular}[m]{c l c l}
        \centering\fcolorbox{black}{color1}{\rule{0pt}{0.7pt}\rule{0.7pt}{0pt}} & \scriptsize Smartphones & \centering\fcolorbox{black}{color6}{\rule{0pt}{0.7pt}\rule{0.7pt}{0pt}} & \scriptsize PCs \\
        \centering\fcolorbox{black}{color3}{\rule{0pt}{0.7pt}\rule{0.7pt}{0pt}} & \scriptsize Home Sensors & \centering\fcolorbox{black}{color7}{\rule{0pt}{0.7pt}\rule{0.7pt}{0pt}} & \scriptsize Printers \\ \centering\fcolorbox{black}{color8}{\rule{0pt}{0.7pt}\rule{0.7pt}{0pt}} & \scriptsize Smart Fitness &
        \centering\fcolorbox{black}{color12}{\rule{0pt}{0.7pt}\rule{0.7pt}{0pt}} & \scriptsize Tablets \\ \centering\fcolorbox{black}{color12}{\rule{0pt}{0.7pt}\rule{0.7pt}{0pt}} & \scriptsize Transportation & \centering\fcolorbox{black}{color13}{\rule{0pt}{0.7pt}\rule{0.7pt}{0pt}} & \scriptsize Smartwatch \\
        \centering\fcolorbox{black}{color4}{\rule{0pt}{0.7pt}\rule{0.7pt}{0pt}} & \scriptsize Indicator & \centering\fcolorbox{black}{color11}{\rule{0pt}{0.7pt}\rule{0.7pt}{0pt}} & \scriptsize Alarms \\
        \centering\fcolorbox{black}{color10}{\rule{0pt}{0.7pt}\rule{0.7pt}{0pt}} & \scriptsize Street Light & \centering\fcolorbox{black}{color5}{\rule{0pt}{0.7pt}\rule{0.7pt}{0pt}} & \scriptsize Parking \\
        \centering\fcolorbox{black}{color9}{\rule{0pt}{0.7pt}\rule{0.7pt}{0pt}} & \scriptsize Car \\
    \vspace{0.2cm}
    \end{tabular}\vspace{-0.6cm}
    \caption{Devices frequency among communities detected in SFOR and SOR relations using Louvain method.}
\label{fig:communities}
\end{figure*}

In Fig.~\ref{fig:communities}, we visualize the detected communities and their sizes for SFOR and SOR. Each number in the x-axis indicates the community label except the last label which indicates others to capture the devices within a community that has less than four devices. We notice that the SFOR relations generate very diversified communities enclosing devices of different types, while the SOR relations, are characterized with  small-size communities formed mainly by a similar type of devices due to the imposed condition that counts devices that are exposed to each other for 30 minutes and more.

\begin{figure}[t]
    \centering
    \includegraphics[width=\columnwidth]{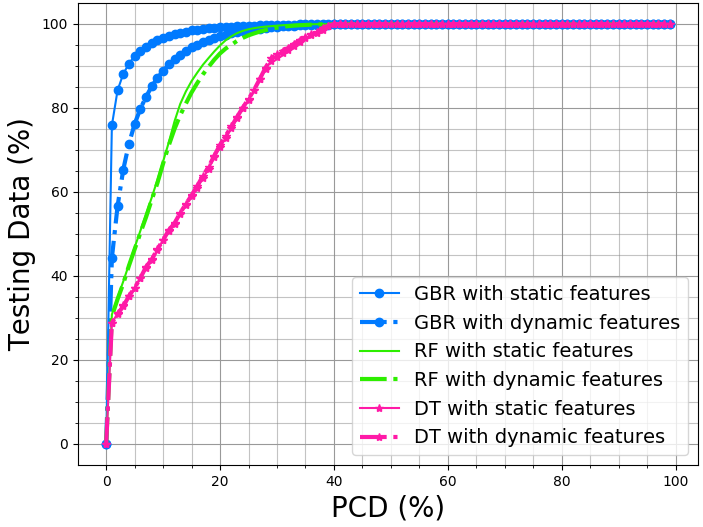}
    \caption{A comparison between the three machine learners for static and dynamic features: cumulative percentage of trained data versus PCD value.}
\label{fig:performance}
\end{figure}

\paragraph{Machine Learning Models}The training history is simulated by creating historical computational sharing activities of the IoT devices. We randomly select combinations of devices (edge computers and requesters) and measure their response times, which can vary for few milliseconds to infinity due to the non-availability of the selected edge computers. Then, we randomly simulate 10.000 sharing experiences between the pairs to train the machine learners to predict the total response time for any given couple of edge computer and requester with respect to their features. Next, the obtained results are pre-processed, i.e., we encode the categorical variables, scale, and normalize the numerical ones. Finally, we pick 75\% and 25\% of  the final pre-processed data set for training and testing, respectively. The models are trained using both dynamic and static features, i.e., we first consider static message sizes for each requester and static availability for each edge computer, e.g., constant operation of the devices, then we consider the case where these features are dynamic and change over time. The process of the hyper-parameter tuning of each model is done using a grid search. The evaluation of our models is mainly based on a custom loss function Percentage
Change Difference (PCD) given as~follows:
\begin{equation}
\label{equ:faza}
PCD = 100 \times \sum\limits_{\substack{i=1}}^{N}{\frac{|y_n - \hat{y}_n|}{
\left(\cfrac{y_n + \hat{y}_n}{2}\right)}},
\end{equation}
where $N$ is the total number of the testing data input, $y_n$ is the
real response time and $\hat{y}_n$ is the predicted response time. We
also use the Mean Squared Error (MSE) and Mean Absolute
Error (MAE) as extra evaluation metrics. The performance of the
models given in Table~\ref{Tab:results} and Fig.~\ref{fig:performance}, shows that the GBR algorithm outperforms all the other models in terms of all the evaluation metrics used. Considering dynamic features slightly affect the models' performance which proves their stability. From Fig.~\ref{fig:performance}, we can notice that more than 85\% of the predictions with GBR have a PCD less than 5\% and 1\% for dynamic and static features, respectively.



\paragraph{Complete Framework Operation}Table~\ref{Tab:results_devices} contains examples of the final outputs of the proposed framework. For example, the first requester is demanding to execute a task of IC=37 and a message size of 1.18 Mb. The TCDP provides 120 potential trustworthy devices having strong relationships with the request and to be fed to the MDP to determine the one offering the lowest response time (RT). The obtained edge computer with ID 6788 provides a RT of 1.12 seconds and has edges of weights 0.25 and 0.5 edges for SFOR and SOR, respectively with the requester.

\begin{table}[]
\caption{Achieved metrics for the different machine learners}\vspace{-0.1cm}
\centering
\begin{tabular}{l c c c}
\toprule
\textbf{Model} & \textbf{MSE}   & \textbf{MAE}                        & \textbf{PCD} \\ \midrule
Gradient Boosting & \textbf{0.002} &\textbf{0.020} & \textbf{1.413}             \\ 
Random Forest     & 0.019 & 0.102                      & 7.062             \\ 
Decision Tree     & 0.062 & 0.185 & 12.444            \\ \bottomrule
\end{tabular}
\label{Tab:results}
\end{table}

\begin{table}[]
\caption{Examples of the framework's outputs}\vspace{-0.1cm}
\centering
\resizebox{\columnwidth}{!}{\begin{tabular}{ >{\centering\arraybackslash}p{1.2cm} >{\centering\arraybackslash}p{0.9cm} c >{\centering\arraybackslash}p{1.1cm} >{\centering\arraybackslash}p{0.8cm} c}
\toprule
\textbf{Size(Mb)/IC} & \textbf{TCDP} & \textbf{MDP (R, RAM)}    & \textbf{Edge ID} &  \textbf{RT (s)}  & \textbf{SFOR/SOR}\\ 
\midrule
1.18 / 37  & 120  & Tablet (1.6 GHz, 8 Go) & 6788  & 1.12  & 0.25 / 0.5 \\ 
5 / 189  & 85   & PC (2.4 GHz, 8 Go) & 1696  & 1.08  & 1 / -  \\ 
0.8 / 23 & 186  & Car (1.2 GHz, 4 Go)&  29  & 1.22  & 0.51 / - \\
\bottomrule
\end{tabular}}
\label{Tab:results_devices}
\end{table}

\section{Conclusion}
In this study, we demonstrated the capability of combining community detection and machine learning in SIoT to enable effective service discovery in determining and assigning edge computers to devices lacking computational resources. The proposed solution reduces the complexity of the service discovery task by shrinking the search space and applying machine learning techniques that do not require recurrent training, which can be very beneficial for large-scale IoT systems.

\vspace{-0.3cm}
\bibliography{references}

\newcommand{\noop}[1]{}
\begin{thebibliography}{10}

\bibitem{cicirelli2017edge}
{F. Cicirelli et al.}, ``Edge computing and social internet of things for
  large-scale smart environments development,'' {\em IEEE Internet of Things
  Journal}, vol.~5, pp.~2557--2571, Nov. 2017.

\bibitem{yu2017survey}
{Yu et al.}, ``A survey on the edge computing for the internet of things,''
  {\em IEEE Access}, vol.~6, pp.~6900--6919, 2017.

\bibitem{zhang2018mobile}
{K. Zhang et al.}, ``Mobile edge computing and networking for green and
  low-latency internet of things,'' {\em IEEE Communications Magazine},
  vol.~56, no.~5, pp.~39--45, 2018.

\bibitem{marche2018dataset}
{C. Marche et al.}, ``A dataset for performance analysis of the social internet
  of things,'' in {\em IEEE Annual International Symposium on Personal, Indoor
  and Mobile Radio Communications (PIMRC'18)}, Bologna, Italy, Sept. 2018.

\bibitem{khanfor2020trustworthy}
{A. Khanfor et al.}, ``A trustworthy recruitment process for spatial mobile
  crowdsourcing in large-scale social iot,'' in {\em IEEE Technology
  Engineering Management Society International Conference (TEMSCON'20),
  Detroit, MI, USA}, Jun. 2020.

\bibitem{alsaffar2016architecture}
{A. Alsaffar et al.}, ``An architecture of {IoT} service delegation and
  resource allocation based on collaboration between fog and cloud computing,''
  {\em Mobile Information Systems}, vol.~2016, Dec. 2016.

\bibitem{samie2016computation}
{F. Farzad et al.}, ``Computation offloading and resource allocation for
  low-power {IoT} edge devices,'' in {\em IEEE 3rd World Forum on Internet of
  Things (WF-IoT'16)}, Dec. Reston, VA, USA, Dec. 2016.

\bibitem{renner2016towards}
T.~Renner, M.~Meldau, and A.~Kliem, ``Towards container-based resource
  management for the internet of things,'' in {\em IEEE International
  Conference on Software Networking (ICSN'16)}, Beijing, China, June 2016.

\bibitem{khanfor2019application}
{A. Khanfor et al.}, ``Application of community detection algorithms on social
  internet-of-things networks,'' in {\em IEEE International Conference on
  Microelectronics (ICM'19), Cairo, Egypt}, pp.~94--97, Dec. 2019.

\bibitem{blondel2008fast}
{Blondel et al.}, ``Fast unfolding of communities in large networks,'' {\em
  Journal of statistical mechanics: theory and experiment}, vol.~2008,
  p.~P10008, Oct. 2008.

\bibitem{10.1023/A:1010933404324}
L.~Breiman, ``Random forests,'' {\em Mach. Learn.}, vol.~45, p.~5–32, Oct.
  2001.

\bibitem{10.5555/1162264}
C.~M. Bishop, {\em Pattern Recognition and Machine Learning (Information
  Science and Statistics)}.
\newblock Berlin, Heidelberg: Springer-Verlag, 2006.

\bibitem{article}
W.-Y. Loh, ``Classification and regression trees,'' {\em Wiley
  Interdisciplinary Reviews: Data Mining and Knowledge Discovery}, vol.~1,
  pp.~14 -- 23, 01 2011.

\bibitem{watts1998collective}
D.~J. Watts and S.~H. Strogatz, ``Collective dynamics of
  ‘small-world’networks,'' {\em nature}, vol.~393, no.~6684, p.~440, 1998.

\end{thebibliography}
\bibliographystyle{ieeetr}
\balance

\end{document}